\documentclass[epjCONF]{svjour}

\usepackage{amsmath}
\usepackage{graphicx}
\usepackage[varg]{txfonts} 
\usepackage[latin1]{inputenc}
\usepackage[mathscr]{eucal}

\session-title{$\,\,\,$}

\begin{document}

\title{The method of unitary clothing transformations in the theory of
nucleon--nucleon scattering }

\author{
I. Dubovyk \inst{1}
\fnmsep\thanks{\email{e.a.dubovyk@mail.ru}}
\and
A. Shebeko \inst{2}
\fnmsep\thanks{\email{shebeko@kipt.kharkov.ua}} }

\institute{Institute of Electrophysics \& Radiation Technologies, \\
              NAS of Ukraine, Kharkov, Ukraine
\and
NSC Kharkov Institute of Physics \& Technology \\
              NAS of Ukraine, Kharkov, Ukraine }

\abstract{ The clothing procedure, put forward in quantum field
theory (QFT) by Greenberg and Schweber, is applied for the
description of nucleon--nucleon ($N$--$N$) scattering. We consider
pseudoscalar ($\pi$ and $\eta$), vector ($\rho$ and $\omega$) and
scalar ($\delta$ and $\sigma$) meson fields interacting with
$1/2$ spin ($N$ and $\bar N$) fermion ones via the
Yukawa--type couplings to introduce trial interactions between
"bare" particles. The subsequent unitary clothing transformations
(UCTs) are found to express the total Hamiltonian through new
interaction operators that refer to particles with physical
(observable) properties, the so--called clothed particles. In this
work, we are focused upon the Hermitian and energy--independent
operators for the clothed nucleons, being built up in the second
order in the coupling constants. The corresponding analytic
expressions in momentum space are compared with the separate meson
contributions to the one--boson--exchange potentials in the meson
theory of nuclear forces. In order to evaluate the $T$ matrix of
the $N$--$N$ scattering we have used an equivalence theorem that
enables us to operate in the clothed particle representation (CPR)
instead of the bare particle representation (BPR) with its huge
amount of virtual processes. We have derived the
Lippmann--Schwinger(LS)--type equation for the CPR elements of the
$T$--matrix for a given collision energy in the two--nucleon
sector of the Hilbert space $\mathcal{H}$ of hadronic states and
elaborated a code for its numerical solution in momentum space.}

\maketitle

\section{Introductory remarks}
\label{UCT_intro}

We know that there are a number of high precision, boson--exchange
models of the two nucleon force $V_{NN}$, such as Paris~ \cite{Lacom80},
Bonn~ \cite{MachHolElst87}, Nijmegen~
\cite{Stocks94}, Argonne~ \cite{WirStocksSchia95}, CD Bonn~
\cite{Mach01} potentials and a fresh family of covariant
one--boson--exchange (OBE) ones~ \cite{GrossStad08}. Note also
successful treatments based on chiral effective field theory~
\cite{OrdoRayKolck94,EpelGloeMeiss00}, for a review see~
\cite{Epel05}.

In this talk, we would like to draw attention to the first
application of unitary clothing transformations
\cite{SheShi01,KorCanShe07} in describing the nucleon-nucleon
($N$--$N$) scattering. Recall that such transformations $W$, being
aimed at the inclusion of the so--called cloud or persistent
effects, make it possible the transition from the bare--particle
representation (BPR) to the clothed--particle representation (CPR)
in the Hilbert space  $\mathcal{H}$ of meson--nucleon states. In
this way, a large amount of virtual processes induced with the
meson absorption/emission, the $N\overline{N}$--pair
annihilation/production and other cloud effects can be accumulated
in the creation (destruction) operators $\alpha_{c}$ for the
"clothed" (physical) me\-sons and nucleons. Such a bootstrap
reflects the most significant distinction between the concepts of
clothed and bare particles.

In the course of the clothing procedure all the generators of the
Poincar\'{e} group get one and the same sparse structure on
$\mathcal{H}$ \cite{SheShi01}. Here we will focus upon one of
them, viz., the total Hamiltonian
\begin{equation}
H = H_{F}(\alpha) + H_{I}(\alpha) \equiv H(\alpha)
\label{UCT_Eq.1}
\end{equation}
with
\begin{equation}
H_{I}(\alpha) = V(\alpha) + \mbox{mass and vertex counterterms},
\label{UCT_Eq.2}
\end{equation}
where free part $H_{F}(\alpha) $  and interaction $V(\alpha)$ depend on
creation (destruction) operators $\alpha^{\dag} (\alpha)$ in the
BPR , i.e., referred to bare particles with physical masses \cite{KorCanShe07}, where they have been introduced via the mass--changing Bogoliu\-bov--type UTs.
To be more definite, let us consider fermions (nucleons and
antinucleons) and bosons ($\pi$--, $\eta$--, $\rho$--,
$\omega$--mesons, etc.) interacting via the Yukawa-type couplings
for scalar (s), pseudoscalar (ps) and vector (v) mesons (see,
e.g., \cite{MachHolElst87}). Then, using a trick prompted by the
derivation of Eq. (7.5.22) in \cite{WeinbergBook1995} to eliminate in a proper way the vector--field component $\varphi_{\rm{v}}^{0}$ , we have
 $V(\alpha)=V_s + V_{ps} + V_{\rm{v}}$ with
\begin{equation}
V_s  = g_s \int d \vec x \, \bar \psi (\vec x) \psi (\vec x)
\varphi_s (\vec x)
\label{UCT_Eq.3}
\end{equation}
\begin{equation}
V_{ps}  = ig_{ps} \int d \vec x \, \bar \psi (\vec x)
\gamma _5 \psi (\vec x) \varphi_{ps} (\vec x)
\label{UCT_Eq.4}
\end{equation}
\begin{multline}
V_{\rm{v}}  =
\int d \vec x \, \left\{ g_{\rm{v}} \bar \psi (\vec x) \gamma_\mu  \psi (\vec x) \varphi_{\rm{v}}^\mu (\vec x) \phantom{\frac{f}{4}} \right. \\
\shoveright{ \left. + \frac{f_{\rm{v}}}{4m} \bar \psi (\vec x) \sigma_{\mu \nu} \psi (\vec x) \varphi_{\rm{v}}^{\mu \nu} (\vec x) \right\} } \\
\shoveleft{ + \int d \vec x \, \left\{ \frac{g_{\rm{v}}^{2}}{2m_{\rm{v}}^2} \bar \psi (\vec x) \gamma_0 \psi (\vec x) \bar \psi (\vec x) \gamma_0 \psi (\vec x) \right. } \\
\left. + \frac{f_{\rm{v}}^{2}}{4m^2} \bar \psi (\vec x)
\sigma_{0i} \psi (\vec x) \bar \psi (\vec x) \sigma_{0i} \psi (\vec x) \right\},
\label{UCT_Eq.5}
\end{multline}
with the boson fields $\varphi_{b}$ and the fermion field $\psi$,
where $ \varphi_{\rm{v}}^{\mu \nu } (\vec x) = \partial^\mu
\varphi_{\rm{v}}^\nu (\vec x) - \partial^\nu \varphi_{\rm{v}}^\mu
(\vec x) $ the tensor of the vector field included.  The mass (vertex) counterterms are given by Eqs. (32)--(33) of Ref. \cite{KorCanShe07} (the difference
$V_{0}(\alpha)$ - $V(\alpha)$ where a primary interaction $V_{0}(\alpha)$ is derived from $V(\alpha)$ replacing the "physical" coupling constants by "bare" ones).

The corresponding set $\alpha$ involves operators $ a^{\dag}(a) $
for the bosons, $ b^{\dag}(b) $ for the nucleons and $ d^{\dag}(d)
$ for the antinucleons. In their terms, e.g., we have the free
pion and fermion parts
\begin{multline}
H_F (\alpha) = \int d \vec k \, \omega_{\vec k} a^\dag  (\vec k) a (\vec k) \\
+ \int d \vec p \, E_{\vec p} \sum \limits_{\mu} \left[ b^\dag
(\vec p, \mu) b (\vec p, \mu) + d^\dag (\vec p, \mu) d (\vec p, \mu) \right]
\label{UCT_Eq.6}
\end{multline}
and the primary trilinear interaction
\begin{equation}
V(\alpha) \sim a b^{\dag} b + a b^{\dag} d^{\dag} + a d b + a d
d^{\dag} + H.c. \label{UCT_Eq.7}
\end{equation}
with the three-legs vertices. Here $\omega_{\vec k}  = \sqrt
{m^2_{b} + \vec k^2 } \,\,\,$($E_{\vec p} = \sqrt {m^2  + \vec
p^2}$) the pion (nucleon) energy with physical mass $m_{b} (m)$,
$\mu$ the fermion polarization index.

We have tried to draw parallels with that field--theoretic
background which has been employed in  boson--ex\-change models.
First of all, we imply the approach by the Bonn group
\cite{MachHolElst87,Mach01}, where, following the idea by
Sch\"{u}tte \cite{Schutte}, the authors started from the total
Hamiltonian (in our notations),
\begin{equation}
H = H_{F}(\alpha) + V(\alpha)
\label{UCT_Eq.8}
\end{equation}
with the boson-nucleon interaction
\begin{equation}
V(\alpha) \sim a b^{\dag} b + H.c.
\label{UCT_Eq.9}
\end{equation}

\section{Analytic expressions for the quasipotentials in momentum space}
\label{UCT_Quasipotential}

As shown in \cite{SheShi01}, after eliminating the so-called bad
terms\footnote{By definition, they prevent the bare vacuum
$\Omega_{0}$ ($a | \Omega_0 \rangle = b | \Omega_0 \rangle
=\ldots= 0$) and the bare one--particle states $|1bare\rangle
\equiv a^{\dag} | \Omega_0 \rangle$ ($b^{\dag} | \Omega_0
\rangle,\ldots$) to be $H$ eigenstates.} from $V(\alpha)$ the
primary Hamiltonian $H(\alpha)$ can be represented in the form,
\begin{equation}
H(\alpha ) = K_F (\alpha _c ) + K_I (\alpha _c ) \equiv K(\alpha_c)
\label{UCT_Eq.10}
\end{equation}
The free part of the new decomposition is determined by
\begin{multline}
K_F(\alpha_c ) = \int d \vec k \, \omega_{\vec k} a_c^\dag (\vec k) a_c (\vec k) \\
+ \int d \vec p \, E_{\vec p} \sum\limits_{\mu}
\left[ b_c^\dag(\vec p, \mu) b_c (\vec p, \mu)
+ d_c^\dag (\vec p, \mu) d_c (\vec p, \mu) \right]
\label{UCT_Eq.11}
\end{multline}
while $K_{I}$ contains only interactions responsible for physical
processes, these quasipotentials between the clothed particles,
e.g.,
\begin{multline}
K_I^{(2)} (\alpha _c ) = K(NN \to NN) + K(\bar N\bar N \to \bar N\bar N) \\
+ K(N\bar N \to N\bar N) + K(b N \to b N) + K(b \bar N \to b \bar N) \\
+ K(b b \to N \bar N) + K(N \bar N \to b b)
\label{UCT_Eq.12}
\end{multline}
In accordance with the clothing procedure developed in~
\cite{SheShi01} they obey the following requirements:

i) The physical vacuum (the $H$ lowest eigenstate) must coincide
with a new no--particle state $\Omega $, i.e., the state that
obeys the equations
\begin{displaymath}
a_c (\vec k) \left\vert \Omega \right\rangle = b_c (\vec p, \mu)
\left\vert \Omega \right\rangle = d_c (\vec p, \mu) \left\vert
\Omega \right\rangle =0,
\end{displaymath}
\begin{equation}
\forall \,\, {\vec k,\,\vec p,\,\mu}
\,\,\,\,\, \left( \left\langle \Omega  | \Omega \right\rangle =1 \right).
\label{UCT_Eq.13}
\end{equation}
ii) New one-clothed-particle states $| \vec k \rangle_c \equiv
a_c^{\dag} (\vec k) \Omega $ etc. are the eigenvectors both of
$K_F$ and $K$,
\begin{equation}
K(\alpha_c)|\vec k \rangle_c=K_F(\alpha_c)|\vec k \rangle_c =
\omega_k|\vec k \rangle_c
\label{UCT_Eq.14}
\end{equation}
\begin{equation}
K_I(\alpha_c)|\vec k \rangle_c=0
\label{UCT_Eq.15}
\end{equation}

iii) The spectrum of indices that enumerate the new operators must
be the same as that for the bare ones .

iv) The new operators $\alpha_{c}$ satisfy the same commutation
rules as do their bare counterparts $\alpha$, since the both sets
are connected to each other via the similarity transformation
\begin{equation}
\alpha_{c}=W^{\dag} \alpha W ,
\label{UCT_Eq.16}
\end{equation}
with a unitary operator $W$ to be obtained as in \cite{SheShi01}.

It is important to realize that operator $K(\alpha_{c})$ is the
same Hamiltonian $H(\alpha)$. Accordingly [10,11] the  $N$--$N$
interaction operator in the CPR has the following structure:
\begin{equation*}
K(NN \rightarrow NN) = \sum\limits_b K_{b}(NN \rightarrow NN),
\end{equation*}
\begin{multline}
K_{b}(NN \rightarrow NN) = \int \sum \limits_\mu d \vec p'_1 \, d \vec p'_2 \, d \vec p_1 \, d \vec p_2 \\
\times V_b (1', 2' ;1, 2) b_c^\dag (1') b_c^\dag (2') b_c (1) b_c(2),
\label{UCT_Eq.17}
\end{multline}
where the symbol $\sum\limits_{\mu}$ denotes the summation over
nucleon spin projections, $1=\{ \vec p_1, \mu_1 \}$, etc.

For our evaluations of the c--number matrices $V_{b}$ we have
employed some experience from Refs. \cite{SheShi01,KorCanShe07} to
get in the second order in the coupling constants
\begin{multline}
V_b (1',2';1,2) = \frac{1}{(2\pi )^3} \frac{m^2}{\sqrt{E_{\vec p'_1} E_{\vec p'_2} E_{\vec p_1} E_{\vec p_2}}} \\
\times \delta \left( \vec p'_1 + \vec p'_2  - \vec p_1  - \vec p_2 \right)
v_b (1',2' ;1,2),
\label{UCT_Eq.18}
\end{multline}
\begin{multline}
v_s (1',2' ;1,2) \\
=  - \frac{g_s^2}{2} \bar u(\vec p'_1) u(\vec p_1 ) \frac{1}{(p_1-p'_1 )^2-m_s^2}
\bar u(\vec p'_2) u(\vec p_2),
\label{UCT_Eq.19}
\end{multline}
\begin{multline}
v_{ps} (1',2' ;1,2) \\
= \frac{g_{ps}^2}{2}\bar u(\vec p'_1) \gamma_5 u(\vec p_1)
\frac{1}{(p_1 - p'_1 )^2 - m_{ps}^2} \bar u(\vec p'_2) \gamma_5
u(\vec p_2 ),
\label{UCT_Eq.20}
\end{multline}
\begin{multline}
v_{\rm{v}} (1',2' ;1,2) =  \frac12 \frac{1}{(p'_1 - p_1 )^2  - m_{\rm{v}}^2} \\
\shoveleft{ \times \left[ \bar u (\vec p'_1) \left\{ ( g_{\rm{v}} + f_{\rm{v}} ) \gamma_\nu - \frac{f_{\rm{v}}}{2m}(p'_1 + p_1)_\nu \right\} u(\vec p_1) \right.}  \\
\shoveright{ \times \bar u (\vec p'_2) \left\{ ( g_{\rm{v}} + f_{\rm{v}} ) \gamma^\nu - \frac{f_{\rm{v}}}{2m}(p'_2 + p_2)^\nu \right\} u(\vec p_2) } \\
\shoveleft{ - \bar u (\vec p'_1) \left\{ (g_{\rm{v}} + f_{\rm{v}}) \gamma_\nu - \frac{f_{\rm{v}}}{2m}(p'_1  + p_1)_\nu \right\} u(\vec p_1) } \\
\shoveright{\times \bar u (\vec p'_2) \frac{f_{\rm{v}}}{2m} \left\{ (\hat p_1^{\prime}+\hat p_2^{\prime}-\hat p_1-\hat p_2)\gamma ^\nu  \right.} \\
\left. \phantom{\frac{f_{\rm{v}}}{2m}} \left.
-(p'_1+p'_2-p_1-p_2)^{\nu} \right\} u( \vec p_2) \right],
\label{UCT_Eq.21}
\end{multline}
where $m_{b}$ the mass of the clothed boson (its physical value)
and $\hat q = q_{\mu}\gamma^{\mu}$. In the framework of the
isospin formalism one needs to add the factor $\vec \tau (1) \vec
\tau (2)$ in the corresponding expressions.

At this point, our derivation of the vector-boson contribution
\eqref{UCT_Eq.21} is to be specifically commented. Actually, it is the
case, where for a Lorentz--invariant Lagrangian it is not
necessarily to have "\dots~the interaction Hamiltonian as the
integral over space of a scalar interaction density; we also need
to add non--scalar terms to the interaction density~\dots" (quoted
from p.292 of Ref. \cite{WeinbergBook1995}). Let us recall that
the density in question has the property,
\begin{equation}
U_F(\Lambda, a) \mathscr{H}(x) U_F^{-1}(\Lambda, a) =
\mathscr{H}(\Lambda x + a)  , \label{UCT_Eq.22}
\end{equation}
where the operators $U_F(\Lambda, a)$ realize a unitary
irreducible representation of the Poincar\'{e} group in the
Hilbert space of states for free (non--interacting) fields.

By definition, the first clothing transformation
$W^{(1)}=\exp[R^{(1)}]$ (${R^{(1)}}^{\dag}=-R^{(1)}$) eliminates
all interactions linear in the coupling constants, viz.,
\begin{equation*}
V^{(1)} = V_s + V_{ps} + V_{\rm{v}}^{(1)},
\end{equation*}
with
\begin{multline}
V_{\rm{v}}^{(1)} = \int d \vec x \left\{ g_{\rm{v}} \bar \psi(\vec
x) \gamma_{\mu} \psi(\vec x) \varphi_{\rm{v}}^{\mu}(\vec x)
\phantom{\frac{f_{\rm{v}}}{4m}} \right. \\
\left. +\frac{f_{\rm{v}}}{4m}\bar \psi(\vec x) \sigma_{\mu \nu} \psi(\vec x) \varphi_{\rm{v}}^{\mu \nu}(\vec x) \right\}.
\label{UCT_Eq.23}
\end{multline}
Following Ref.\cite{SheShi01} we have
\begin{equation}
R^{(1)} = -i\lim \limits_{\varepsilon \rightarrow 0+} \int \limits_0^{\infty} V_D^{(1)}(t)e^{-\varepsilon t} dt
\label{UCT_Eq.24}
\end{equation}
if $m_b < 2m$. Here
\begin{equation*}
V_D^{(1)} \equiv \exp[iH_Ft] V^{(1)} \exp[-iH_Ft] = \int
\mathscr{H}^{(1)} (x) d \vec x,
\end{equation*}
where $\mathscr{H}^{(1)} (x)$ is the Lorentz scalar.

The corresponding interaction operator in the CPR
\eqref{UCT_Eq.12} can be written as
\begin{equation}
K_I^{(2)}(\alpha_c) = \frac{1}{2}\left[ R^{(1)}(\alpha_c),
V^{(1)}(\alpha_c) \right] + V^{(2)}(\alpha_c),
\label{UCT_Eq.25}
\end{equation}
where we have kept only the contributions of the second order in
the coupling constants, so
\begin{multline}
V^{(2)} = \int d \vec x \left\{  \frac{g_{\rm{v}}^{2}}{2m_{\rm{v}}^2}\bar \psi(x\vec ) \gamma_0 \psi(\vec x) \bar \psi(\vec x) \gamma_0 \psi(\vec x) \right. \\
\left. +\frac{f_{\rm{v}}^{2}}{4m^2}\bar \psi(\vec x) \sigma_{0i}
\psi(\vec x)\bar \psi(\vec x) \sigma_{0i} \psi(\vec x)  \right\}.
\label{UCT_Eq.26}
\end{multline}
We point out that all quantities in the r.h.s. of Eq.\eqref{UCT_Eq.25}
depend on the new creation(destruction) operators $\alpha_c$. In
particular, it means that in the standard Fourier expansions of
the fields involved in the definitions of $V^{(1)}$ and $V^{(2)}$
one should replace the set $\{ \alpha \}$ by the set $\{ \alpha_c
\}$. Thus, there is an essential distinction between
$V^{(1)}$($V^{(2)}$), on the one hand, and the first(second)
integral in the r.h.s. of Eq.\eqref{UCT_Eq.5}, on the other hand.

For this exposition we do not intend to derive all interactions
between the clothed mesons and nucleons, allowed by formula
\eqref{UCT_Eq.25}. Our aim is more humble, viz., to find in the r.h.s.
of Eq.\eqref{UCT_Eq.25} terms of the type \eqref{UCT_Eq.17}, responsible
for the $N$--$N$ interaction. Meanwhile, in case of the vector mesons
we encounter an interplay between the commutator
$[R^{(1)}_{\rm{v}},V_{\rm{v}}^{(1)}]/2$ and the integral
\eqref{UCT_Eq.26}. Indeed, after a simple algebra we find
\begin{multline*}
\frac12 \left[ R^{(1)}, V^{(1)} \right]_{\rm{v}}(NN \rightarrow NN) \\
= K_{\rm{v}}(NN \rightarrow NN) + K_{cont}(NN \rightarrow NN),
\end{multline*}
where the first term has the structure of Eq.\eqref{UCT_Eq.17} with
the coefficients by \eqref{UCT_Eq.21}. At the same time the second
term $K_{cont}$ completely cancels the
non--scalar operator $V^{(2)}$. The latter may be associated with
a contact interaction since it does not contain any propagators
(cf. the approach by the Osaka group \cite{TamuraSato88}), being
expressed through the $b^{\dag}_c(b_c)$. In other words, the first
UCT enables us to remove the non--invariant terms directly in the
Hamiltonian. In our opinion, such a cancellation, first discussed
here, is a pleasant feature of the CPR.

\begin{figure}[h]
  \includegraphics[scale=0.55]{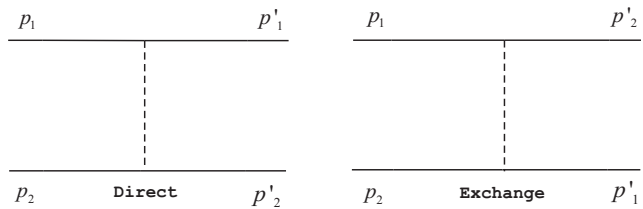}

\caption{The Feynman--like diagrams for the direct and exchange
contributions in the r.h.s. of Eq.\eqref{UCT_Eq.27}.} \label{UCT_fig:1}
\end{figure}

Moreover, as it was shown in Ref.\cite{SheShi01}, for each boson
included the corresponding relativistic and properly symmetrized
$N$--$N$ interaction, the kernel of integral equations for the $N$--$N$
bound and scattering states, is determined by
\begin{multline}
\left\langle b_c^\dag  (\vec p'_1) b_c^\dag (\vec p'_2) \Omega
\right| K_b(NN \rightarrow NN)
\left| b_c^\dag (\vec p_1) b_c^\dag (\vec p_2) \Omega \right\rangle \\
= V_b^{dir}(1',2';1,2) - V_b^{exc}(1',2';1,2),
\label{UCT_Eq.27}
\end{multline}
where we have separated the so--called direct
\begin{equation}
V_b^{dir} (1',2' ;1,2) =  - V_b (1',2' ;1,2) - V_b (2',1' ;2,1)
\label{UCT_Eq.28}
\end{equation}
and exchange
\begin{equation}
V_b^{exc} (1',2' ;1,2) = V_b^{dir} (2',1' ;1,2)
\label{UCT_Eq.29}
\end{equation}
terms. For example, the one--pion--exchange contribution can be
divided into the two parts:
\begin{multline}
V_{\pi}^{dir}(1',2';1,2) = - \frac{g_{\pi}^2}{(2\pi )^3} \frac{m^2}{\sqrt {E_{\vec p'_1} E_{\vec p'_2} E_{\vec p_1} E_{\vec p_2}}} \\
\times \delta \left( \vec p'_1  + \vec p'_2  - \vec p_1  - \vec p_2 \right) \bar u(\vec p'_1) \gamma_5 u(\vec p_1) \bar u(\vec p'_2) \gamma_5 u(\vec p_2) \\
\times \frac12 \left\{ \frac{1}{(p_1 - p'_1 )^2 - m_{\pi}^2} +
\frac{1}{(p_2 - p'_2 )^2 - m_{\pi}^2} \right\}
\label{UCT_Eq.30}
\end{multline}
and
\begin{multline}
V_{\pi}^{exc}(1',2';1,2) = - \frac{g_{\pi}^2}{(2\pi )^3} \frac{m^2}{\sqrt {E_{\vec p'_1} E_{\vec p'_2} E_{\vec p_1} E_{\vec p_2}}} \\
\times \delta \left( \vec p'_1  + \vec p'_2  - \vec p_1  - \vec p_2 \right) \bar u(\vec p'_1) \gamma_5 u(\vec p_2) \bar u(\vec p'_2) \gamma_5 u(\vec p_1) \\
\times \frac12 \left\{ \frac{1}{(p_2 - p'_1 )^2 - m_{\pi}^2} +
\frac{1}{(p_1 - p'_2 )^2 - m_{\pi}^2} \right\}
\label{UCT_Eq.31}
\end{multline}
to be depicted in Fig.\ref{UCT_fig:1}, where the dashed lines correspond to the
following Feynman--like "propagators":
\begin{displaymath}
\frac12 \left\{ \frac{1}{(p_1 - p'_1 )^2 - m_{\pi}^2} +
\frac{1}{(p_2 - p'_2 )^2 - m_{\pi}^2} \right\}
\end{displaymath}
on the left panel and
\begin{displaymath}
\frac12 \left\{ \frac{1}{(p_2 - p'_1 )^2 - m_{\pi}^2} +
\frac{1}{(p_1 - p'_2 )^2 - m_{\pi}^2} \right\}
\end{displaymath}
on the right panel. Other distinctive features of the result
\eqref{UCT_Eq.27} have been discussed in
\cite{SheShi01,KorCanShe07}.

\section{ The field--theoretic description of the elastic
\textsl{N}--\textsl{N} scattering }
\label{UCT_NN_scat}
\subsection{ The $T$--matrix in the CPR }
\label{UCT_T_matrix}
In order to evaluate the $N$--$N$ scattering amplitude for the
collision energy $E$ we will regard a field operator $T$ that
meets the equation
\begin{equation}
T(E+i0)=H_{I}+H_{I}(E+i0-H_F)^{-1}T(E+i0)
\label{UCT_Eq.32}
\end{equation}
and whose matrix elements
$\langle N N |T(E+i0) | N N \rangle$
on the energy shell
$
E=E_1 + E_2 = E_1^{'} + E_2^{'}
$
can be expressed through the phase shifts and mixing parameters.

Unlike nonrelativistic quantum mechanics (NQM) in relativistic
QFT the interaction $H_I$ does not conserve the particle number,
being the spring of particle creation and destruction. The feature
makes the problem of finding  the $N$--$N$
scattering matrix much more complicated than in the framework of nonrelativistic
approach.

Such a general field--theoretic consideration can be simplified
with the help of an equivalence theorem \cite{ShebekoFB17}
according to which the $S$ matrix elements in the Dirac
(\textrm{D}) picture, viz.,
\begin{equation}
S_{fi} \equiv \langle \alpha^{\dag}...\Omega_0 | S(\alpha) |
\alpha^{\dag}...\Omega_0 \rangle
\label{UCT_Eq.33}
\end{equation}
are equal to the corresponding elements
\begin{equation}
S_{fi}^c \equiv \langle \alpha_c^{\dag}...\Omega | S(\alpha_c) |
\alpha_c^{\dag}...\Omega \rangle
\label{UCT_Eq.34}
\end{equation}
of the $S$ matrix in the CPR once the UCTs
$$W_D(t) =
\exp(iK_Ft)W\exp(-iK_Ft)$$
obey the condition
\begin{equation}
W_D(\pm\infty)=1 \label{UCT_Eq.35}
\end{equation}
The $T$ operator in the CPR satisfies the equation
\begin{multline}
T_{cloth}(E+i0)=K_{I} \\
+K_{I}(E+i0-K_F)^{-1}T_{cloth}(E+i0)
\label{UCT_Eq.36}
\end{multline}
and the matrix
\begin{multline}
T_{fi} \equiv \langle f; b | T(E+i0) | i; b \rangle  \\
= \langle f; c| T_{cloth}(E+i0) | i; c \rangle \equiv T_{fi}^{c},
\label{UCT_Eq.37}
\end{multline}
where $| ; b\rangle$ ( $| ; c\rangle$ ) are the $H_F$ ( $K_F$ )
eigenvectors, may be evaluated relying upon properties of the new
interaction $K_I(\alpha_c)$.

If in Eq.\eqref{UCT_Eq.36} we approximate $K_I$
by $K_I^{(2)}$, then initial task of evaluating the CPR matrix
elements  can be reduced to
solving the equation
\begin{multline}
\langle 1', 2' | T_{NN}(E) | 1, 2 \rangle = \langle 1', 2' | K_{NN} | 1, 2 \rangle \\
 +  \langle 1', 2' | K_{NN} (E+i0-K_F)^{-1} T_{NN}(E) | 1, 2 \rangle.  \label{UCT_Eq.38}
\end{multline}

\subsection{ The $R$--matrix equation and its angular--momentum decomposition}
\label{UCT_R_matrix}

For practical applications one prefers to work with the
corresponding $R$--matrix that meets the set of equations
\begin{multline}
\left\langle {1'2'} \right|\bar{R}(E)\left| {12} \right\rangle =
\left\langle {1'2'} \right|\bar{K}_{NN} \left| {12} \right\rangle \\
+ \int\limits_{34} \!\!\!\!\!\!\!\! \sum {\left\langle {1'2'}
\right|\bar{K}_{NN} \left| {34} \right\rangle \frac{{\left\langle
{34} \right|\bar{R}(E)\left| {12} \right\rangle }}{{E - E_3  - E_4}}}
\label{UCT_Eq.39}
\end{multline}
with $\bar{R}(E) = R(E)/2$ and $\bar{K}_{NN} = K_{NN}/2 $, where
the operation $\int\limits_{34} \!\!\!\!\!\! \sum$ means
the summation over nucleon polarizations and the $p.v.$ integration over
nucleon momenta. The kernel of Eq.\eqref{UCT_Eq.39} is
\begin{multline*}
\left\langle {1'2'} \left| \bar K_{NN} \right| {12} \right\rangle =
\delta \left( \vec p'_1 + \vec p'_2 - \vec p_1  - \vec p_2  \right) \left\langle {1'2'} \left| \bar V \right| {12} \right\rangle \\
\shoveleft{\equiv \delta \left( \vec p'_1 + \vec p'_2 - \vec p_1  - \vec p_2  \right)} \\
\times \left\langle {\vec p'_1 \mu'_1 \tau'_1 , \vec p'_2 \mu'_2
\tau'_2 \left| {\bar V} \right| \vec p_1 \mu _1 \tau _1 ,\vec p_2
\mu _2 \tau _2 } \right\rangle
\end{multline*}

The subsequent calculations are essentially
simplified in the center--of--mass system (c.m.s) in which
\begin{multline}
\left\langle \vec p' \mu'_1 \mu'_2, \tau'_1 \tau'_2 \left| {\bar R(E)} \right| \vec p \mu _1 \mu _2, \tau_1 \tau_2  \right\rangle  \\
\shoveright{= \left\langle {\vec p' \mu'_1 \mu'_2, \tau'_1 \tau'_2 \left| {\bar V} \right| \vec p \mu _1 \mu _2, \tau_1 \tau_2 } \right\rangle} \\
\shoveleft{+ \sum p.v. \int d \vec q  \left\langle
\vec p' \mu'_1 \mu'_2, \tau'_1 \tau'_2 \left| {\bar V } \right| \vec q \mu _3 \mu _4, \tau_3 \tau_4 \right\rangle} \\
\times \frac{\left\langle \vec q \mu _3 \mu _4, \tau_3 \tau_4
\left| {\bar R(E)} \right| \vec p \mu _1 \mu _2, \tau_1 \tau_2
\right\rangle}{E - 2E_{\vec q}}
\label{UCT_Eq.40}
\end{multline}
Here the quantum numbers $\mu(\tau)$ are the individual spin
(isospin) projections.

Accordingly Eq. \eqref{UCT_Eq.27}
\begin{multline}
\left\langle {1'2'} \right|{\bar V }\left| {12} \right\rangle =
\frac{1}{2(2\pi )^3} \frac{m^2}{E_{\vec p'} E_{\vec p}} \\
\times \sum\limits_b [v_b^{dir} (1',2';1,2) -
v_b^{exc}(1',2';2,1)]
\label{UCT_Eq.41}
\end{multline}
with
\begin{equation}
v_b^{dir} (1',2';1,2) = - v_b(1',2';1,2) - v_b(2',1';2,1)
\label{UCT_Eq.42}
\end{equation}
and
\begin{equation*}
v_b^{exc}(1',2';2,1) = v_b^{dir} (2',1';1,2),
\end{equation*}
where the separate boson contributions are determined by Eqs.
\eqref{UCT_Eq.19}--\eqref{UCT_Eq.21} with $\vec p_1 = \vec p = - \vec p_2$
and $\vec p'_1 = \vec p' = -\vec p'_2$.

Following a common practice we are interested in the angular--momentum
decomposition of Eq.\eqref{UCT_Eq.40} assuming a
nonrelativistic analog of relativistic partial wave expansions
(see \cite{Werle66} and refs. therein) for two--particle states.
For example, the clothed two--nucleon state (the so--called
two--nucleon plane wave) can be represented as
\begin{multline}
\left| \vec p \mu _1 \mu _2, \tau _1 \tau _2 \right\rangle  =
\sum \left( \tfrac12 \mu_1 \tfrac12 \mu_2 \left| SM_S \right. \right) \left( \tfrac12 \tau _1 \tfrac12 \tau _2 \left| TM_T \right. \right) \\
\left( lm_l SM_S \left| JM_J \right. \right) Y_{lm_l }^* \left(
\vec p /p \right) \left| p J(lS)M_J ,TM_T \right\rangle,
\label{UCT_Eq.43}
\end{multline}
where $J$, $S$ and $T$ are, respectively, total angular momentum,
spin and isospin of the $NN$ pair, being the eigenvalues of the
operators $\vec J_{ferm}$, $\vec{S}_{ferm}$ and $\vec T_{ferm}$.
Here
\begin{equation}
\vec J_{ferm}  = \vec{L}_{ferm}+\vec{S}_{ferm},
\label{UCT_Eq.44}
\end{equation}
where  $\vec{L}_{ferm}$ ($\vec{S}_{ferm}$) the orbital (spin)
momentum of the fermion field,
\begin{multline}
\vec{L}_{ferm}=\frac{i}{2} \sum\limits_\mu  \int  d \vec p \,\,\vec p \times
\left[ \frac{\partial b_c^\dag (\vec p \mu)} {\partial \vec p} b_c(\vec p\mu ) \right. \\
- b_c^\dag  (\vec p\mu) \frac{\partial b_c(\vec p\mu )}{\partial \vec p}
+ \frac{\partial d_c^\dag  (\vec p\mu )}{\partial \vec p}d_c(\vec p\mu ) \\
\left. - d_c^\dag  (\vec p\mu ) \frac{\partial d_c(\vec p\mu )}{\partial \vec p} \right]
\label{UCT_Eq.45}
\end{multline}
and
\begin{multline}
\vec{S}_{ferm} = \frac{1}{2} \sum \limits_{\mu \mu'} \int  d \vec
p \,\,  \chi^{\dag}_{\mu'} \mathbf{\sigma} \chi_{\mu} \left \{
b_c^\dag (\vec p \mu ')b_c(\vec p \mu ) \right. \\
\left. - d_c^\dag (\vec p \mu ')d_c(\vec p \mu ) \right \},
\label{UCT_Eq.46}
\end{multline}
where $\chi_{\mu'}(\chi_\mu)$ are the Pauli spinors. For brevity, we do not show
the isospin operator $\vec T_{ferm}$.

The corresponding eigenvalue equations look as
\begin{equation*}
{\vec{J}}^{\,2}_{ferm} \left| {pJ(lS)M_J } \right\rangle = J(J+1)
\left| {pJ(lS)M_J } \right\rangle
\end{equation*}
\begin{equation}
J_{ferm}^z \left| {pJ(lS)M_J } \right\rangle = M_J \left|
{pJ(lS)M_J } \right\rangle
\label{UCT_Eq.47}
\end{equation}
and
\begin{equation*}
{\vec{S}}^{\,2}_{ferm} \left| {\vec pSM_S } \right\rangle = S(S+1)
\left| {\vec pSM_S } \right\rangle
\end{equation*}
\begin{equation}
S_{ferm}^z \left| {\vec pSM_S } \right\rangle = M_S \left| {\vec
pSM_S } \right\rangle
\label{UCT_Eq.48}
\end{equation}
Doing so, we have introduced the vectors\footnote{For a
moment, the isospin quantum numbers are suppressed.}
\begin{equation}
\left| {\vec pSM_S } \right\rangle  = \sum {\left( { \left.
\frac{1}{2} \mu _1 \frac{1}{2} \mu _2 \right| SM_S } \right)}
\left| {\vec p\mu _1 \mu _2 } \right\rangle
\label{UCT_Eq.49}
\end{equation}
and
\begin{multline}
\left| {pJ(lS)M_J } \right\rangle \\
= \int { d \hat{\vec{p}}  } \,
Y_{lm_l } \left( \vec{p}/p \right) \left| {\vec p SM_S }
\right\rangle \left( {lm_l SM_S \left| {JM_J } \right.} \right)
\label{UCT_Eq.50}
\end{multline}

A simple way of deriving Eqs.\eqref{UCT_Eq.47}--\eqref{UCT_Eq.48} is to use the
transformation
\begin{equation}
U_{F}^{c}(R) | \vec p S M_S \rangle = |R \vec p SM_S^{\prime} \rangle D_{M_S^{\prime}M_S}^{(S)}(R)
\label{UCT_Eq.51}
\end{equation}
\begin{equation*}
\forall \, R \, \in \, \mbox{the rotation group}
\end{equation*}
One should note that in our case the separable ansatz
\begin{equation*}
\left|{\vec p_1 \vec p_2 \mu _1 \mu _2 } \right\rangle  = \left|
{\vec p_1 \mu _1 } \right\rangle \left| {\vec p_2 \mu _2 }
\right\rangle
\end{equation*}
often exploited in relativistic quantum mechanics (RQM) (see,
e.g., \cite{Werle66} and \cite{KeisPoly91}) does not work.
However, one can employ the similarity transformation
\footnote{Sometimes it is convenient to handle the operators
$b_c^{\dag}(p\mu)=\sqrt{p_0}b_c^{\dag}(\vec{p}\mu)$ and their
adjoints $b_c(p\mu)$ that meet covariant relations $\left\{
{b_c^\dag (p'\mu '),b_c(p\mu )} \right\} = p_0 \delta (\vec p' -
\vec p)\delta_{\mu' \mu}$ }
\begin{multline}
U_F^{c} \left( {\Lambda ,a} \right)b_c^\dag  \left( {p\mu }
\right)U_F^{c}\dag \left( {\Lambda ,a} \right) \\
= e^{i\Lambda p \cdot a} b_c^\dag \left( {\Lambda p\mu '} \right) D_{\mu
'\mu}^{(\frac12)} \left( {W\left( {\Lambda ,p} \right)} \right)
\label{UCT_Eq.52}
\end{multline}
with the Wigner rotation $W(\Lambda,p)$ (e.g., for rotations
$W(R,p)=R$ ) and the property of the physical vacuum $\Omega $ to
be invariant with respect to unitary transformations $U_F^{c}$ in the CPR
(some details can be found in a separate paper).

The use of expansion \eqref{UCT_Eq.43} gives rise to the well known JST
representation, in which
\begin{multline}
\left\langle p' J'(l'S')M'_J, T'M'_T \left| \bar R(E)\{ \bar V\} \right| p J(lS)M_J, TM_T  \right\rangle \\
= \bar R(E)\{ \bar V\} _{l'l}^{JST} (p',p) \delta _{J'J} \delta _{M'_J M_J }
\delta _{S'S} \delta _{T'T} \delta _{M'_T M_T }, \label{UCT_Eq.53}
\end{multline}
so Eq.\eqref{UCT_Eq.40} reduces to the set of simple
integral equations,
\begin{multline}
\bar R_{l'\,l}^{JST}(p',p) = \bar V_{l'\,l}^{JST}(p',p) \\
+ \sum\limits_{l''} {p.v.} \int\limits_0^\infty \frac{q^2 \,
dq}{2(E_p  - E_q )} \bar V_{l'\,l''}^{JST}(p',q) \bar R_{l''{\rm{
}}l}^{JST}(q,p) \label{UCT_Eq.54}
\end{multline}
to be solved for each submatrix $\overline{R}^{JST}$ composed of
the elements
\begin{equation}
\bar R_{l'l}^{JST}(p',p) \equiv \bar R_{l'l}^{JST}(p',p;2E_p),
\label{UCT_Eq.55}
\end{equation}
where $E_p = \sqrt{\vec p^2 + m^2}$ the collision energy in the
c.m.s.. One should note that in view of the charge independence
assumed in this work one has to solve two separate equations for
isospin values $T=0$ and $T=1$.

\section{Results of numerical calculations and their discussion}
\label{UCT_numresult}
In the course of our computations we have used the so--called
matrix inversion method (MIM) (see \cite{BJ76} and refs. therein).
Since we deal with the relativistic dispersion law for the
particle energies, the well known substraction procedure within
the MIM leads to equations
\begin{multline}
R_{l'{\rm{ }}l}^{JST} (p',p) = V_{l'{\rm{ }}l}^{JST}(p',p) \\
+ \frac12 \sum \limits_{l''} \int \limits_0^\infty \frac{d q}{p^2
- q^2}
\left\{ q^2 (E_p + E_q) V_{l'l''}^{JST}(p',q) R_{l''l}^{JST}(q,p) \right. \\
\left. - 2p^2 E_p V_{l'l''}^{JST}(p',p) R_{l''l}^{JST}(p,p)
\right\} . \label{UCT_Eq.56}
\end{multline}
\begin{table}[!ht]
\caption{The best--fit parameters for the two models. The third (fourth) column taken from Table A.1 \cite{Mach89} (obtained by solving Eqs.\eqref{UCT_Eq.56} with a least squares fitting to OBEP values in Table 2). All masses
are in $MeV$, and $n_b=1$ except for $n_{\rho}=n_{\omega}=2.$}
\label{UCT_tab:1} \tabcolsep=0.05\columnwidth

\begin{tabular}{cccc}

\hline \hline \noalign{\smallskip}

Meson &  & Potential B & UCT \\

\noalign{\smallskip}\hline\noalign{\smallskip}

$\pi$    & $g^2_{\pi}/4\pi$  & 14.4   & 14.5    \\
         & $\Lambda_{\pi}$   & 1700   & 2200    \\
         & $m_{\pi}$         & 138.03 & 138.03  \\

\noalign{\smallskip}\hline\noalign{\smallskip}

$\eta$    & $g^2_{\eta}/4\pi$  & 3     & 2.8534  \\
          & $\Lambda_{\eta}$   & 1500  & 1200    \\
          & $m_{\eta}$         & 548.8 & 548.8   \\

\noalign{\smallskip}\hline\noalign{\smallskip}

$\rho$    & $g^2_{\rho}/4\pi$    & 0.9  & 1.3   \\
          & $\Lambda_{\rho}$     & 1850 & 1450  \\
          & $f_{\rho}/g_{\rho}$  & 6.1  & 5.85  \\
          & $m_{\rho}$           & 769  & 769   \\

\noalign{\smallskip}\hline\noalign{\smallskip}

$\omega$    & $g^2_{\omega}/4\pi$  & 24.5  & 27      \\
            & $\Lambda_{\omega}$   & 1850  & 2035.59 \\
            & $m_{\omega}$         & 782.6 & 782.6   \\

\noalign{\smallskip}\hline\noalign{\smallskip}

$\delta$    & $g^2_{\delta}/4\pi$  & 2.488 & 1.6947  \\
            & $\Lambda_{\delta}$   & 2000  & 2200    \\
            & $m_{\delta}$         & 983   & 983     \\

\noalign{\smallskip}\hline\noalign{\smallskip}

$\sigma, \,T=0 $    & $g^2_{\sigma}/4\pi$  & 18.3773   & 19.4434   \\
                    & $\Lambda_{\sigma}$   & 2000      & 1538.13   \\
                    & $m_{\sigma}$         & 720       & 717.7167  \\

\noalign{\smallskip}\hline\noalign{\smallskip}

$\sigma, \, T=1 $    & $g^2_{\sigma}/4\pi$  & 8.9437    &  10.8292   \\
                     & $\Lambda_{\sigma}$   & 1900      &  2200      \\
                     & $m_{\sigma}$         & 550       &  568.8612  \\

\noalign{\smallskip}\hline\noalign{\smallskip}

\end{tabular}
\end{table}
To facilitate comparison with some derivations and calculations
from Refs. \cite{MachHolElst87}, \cite{Mach89}, we introduce the
notation
\begin{multline*}
\left\langle {\vec p'\,\mu'_1 \mu' _2 } \right| v^{UCT}_b \left|
{\vec p\,\mu _1 \mu _2 } \right\rangle \\ \equiv
-F^2_b(p',p)
\left[ v_b (1',2';1,2)+v_b (2',1' ;2,1) \right]
\end{multline*}
for the regularized UCT quasipotentials
in the c.m.s. As in Ref.\cite{MachHolElst87}, we put that invariants $F_b(p',p)= F_b(\Lambda p', \Lambda p)$ have a phenomenological form,
\begin{equation*}
F_b(p',p) = \left[ \frac{\Lambda _b^2  - m_{b}^2}{\Lambda _{b}^2
- (p' - p)^2} \right]^{n_{b}} \equiv F_b[(p'-p)^2]
\end{equation*}
Doing so, we have
\begin{multline}
\left\langle \vec p^{\, \prime} \,\mu'_1 \mu' _2  \right|   v^{UCT}_{s} \left| \vec p \,\mu _1 \mu _2 \right\rangle \\
= g_s^2 \bar u( \vec p^{\, \prime}) u(\vec p ) \frac{F^2_{s}[(p'-p)^2]}{(p' - p )^2  - m_s^2} \bar u( - \vec p^{\, \prime} ) u( - \vec p ),
\label{UCT_Eq.57}
\end{multline}
\begin{multline}
\left\langle \vec p^{\, \prime} \,\mu'_1 \mu' _2 \right|   v^{UCT}_{ps} \left| \vec p\,\mu _1 \mu _2 \right\rangle \\
= - g_{ps}^2 \bar u(\vec p^{\, \prime}) \gamma _5 u(\vec p)  \frac{F^2_{ps}[(p'-p)^2]}{(p' - p )^2 - m_{ps}^2} \bar u( - \vec p^{\, \prime}) \gamma _5 u( - \vec p )
\label{UCT_Eq.58}
\end{multline}
\begin{table*}[!ht]
\caption{Neutron--proton phase shifts (in degrees) for various
laboratory energies (in MeV). The OBEP(OBEP$^*$)--rows taken from Table~5.2 \cite{Mach89} (calculated by solving Eqs.\eqref{UCT_Eq.61} with the model parameters from the third column in Table 1). The UCT$^*$(UCT)--rows calculated by solving Eqs.\eqref{UCT_Eq.56} with the parameters from the third (fourth) column in Table 1. As in \cite{MachHolElst87}, we have used the bar convention \cite{Stapp} for the phase parameters.}
\label{UCT_tab:2} \tabcolsep=0.08\columnwidth
\begin{center}
\begin{tabular}{cccccccc}

\hline \hline \noalign{\smallskip}

State & Potential & 25 & 50 & 100 & 150 & 200 & 300 \\

\noalign{\smallskip}\hline\noalign{\smallskip}

            & OBEP     & 50.72 & 39.98 & 25.19 & 14.38 & 5.66 & -8.18  \\
{${^1}S_0$} & OBEP$^*$ & 50.71 & 39.98 & 25.19 & 14.37 & 5.66 & -8.18  \\
            & UCT$^*$  & 66.79 & 53.01 & 36.50 & 25.27 & 16.54 & 3.12  \\
            & UCT      & 50.03 & 39.77 & 25.55 & 15.20 & 6.92 & -6.07  \\

\noalign{\smallskip}\hline\noalign{\smallskip}

            & OBEP     & -7.21 & -11.15 & -16.31 & -20.21 & -23.47 & -28.70  \\
{${^1}P_1$} & OBEP$^*$ & -7.17 & -11.15 & -16.32 & -20.21 & -23.48 & -28.71  \\
            & UCT$^*$  & -7.40 & -11.70 & -17.73 & -22.63 & -26.98 & -34.54  \\
            & UCT      & -7.15 & -10.95 & -15.62 & -18.90 & -21.49 & -25.41  \\

\noalign{\smallskip}\hline\noalign{\smallskip}

            & OBEP     & 0.68 & 1.58 & 3.34 & 4.94 & 6.21 & 7.49  \\
{${^1}D_2$} & OBEP$^*$ & 0.68 & 1.58 & 3.34 & 4.94 & 6.21 & 7.49  \\
            & UCT$^*$  & 0.68 & 1.59 & 3.40 & 5.10 & 6.52 & 8.20  \\
            & UCT      & 0.68 & 1.56 & 3.22 & 4.68 & 5.77 & 6.68  \\

\noalign{\smallskip}\hline\noalign{\smallskip}

            & OBEP     & 9.34 & 12.24 & 9.80  & 4.57 & -1.02 & -11.48   \\
{${^3}P_0$} & OBEP$^*$ & 9.34 & 12.24 & 9.80  & 4.57 & -1.02 & -11.48   \\
            & UCT$^*$  & 9.48 & 12.53 & 10.32 & 5.27 & -0.15 & -10.27   \\
            & UCT      & 9.30 & 12.16 & 9.81  & 4.73 & -0.68 & -10.76   \\

\noalign{\smallskip}\hline\noalign{\smallskip}

            & OBEP     & -5.33 & -8.77 & -13.47 & -17.18 & -20.49 & -26.38  \\
{${^3}P_1$} & OBEP$^*$ & -5.33 & -8.77 & -13.47 & -17.18 & -20.48 & -26.38  \\
            & UCT$^*$  & -5.27 & -8.62 & -13.09 & -16.56 & -19.63 & -25.06  \\
            & UCT      & -5.28 & -8.58 & -12.85 & -16.06 & -18.86 & -23.79  \\

\noalign{\smallskip}\hline\noalign{\smallskip}

            & OBEP     & 3.88 & 9.29 & 17.67 & 22.57 & 24.94 & 25.36  \\
{${^3}D_2$} & OBEP$^*$ & 3.89 & 9.29 & 17.67 & 22.57 & 24.94 & 25.36  \\
            & UCT$^*$  & 3.86 & 9.15 & 17.12 & 21.51 & 23.47 & 23.48  \\
            & UCT      & 3.89 & 9.25 & 17.31 & 21.77 & 23.75 & 23.61  \\

\noalign{\smallskip}\hline\noalign{\smallskip}

            & OBEP     & 80.32 & 62.16 & 41.99 & 28.94 & 19.04 & 4.07   \\
{${^3}S_1$} & OBEP$^*$ & 80.31 & 62.15 & 41.98 & 28.93 & 19.03 & 4.06   \\
            & UCT$^*$  & 92.30 & 72.71 & 51.44 & 38.10 & 28.20 & 13.70  \\
            & UCT      & 79.60 & 61.53 & 41.57 & 28.75 & 19.08 & 4.60  \\

\noalign{\smallskip}\hline\noalign{\smallskip}

            & OBEP     & -2.99 & -6.86 & -12.98 & -17.28 & -20.28 & -23.72  \\
{${^3}D_1$} & OBEP$^*$ & -2.99 & -6.87 & -12.99 & -17.28 & -20.29 & -23.72  \\
            & UCT$^*$  & -2.74 & -6.43 & -12.36 & -16.54 & -19.47 & -22.78  \\
            & UCT      & -3.00 & -6.90 & -13.12 & -17.66 & -21.11 & -26.03  \\

\noalign{\smallskip}\hline\noalign{\smallskip}

                  & OBEP     & 1.76  & 2.00  & 2.24  & 2.58  & 3.03 & 4.03  \\
{$\varepsilon_1$} & OBEP$^*$ & 1.76  & 2.00  & 2.24  & 2.58  & 3.03 & 4.03  \\
                  & UCT$^*$  & 0.02  & -0.12 & -0.17 & 0.04  & 0.41 & 1.40  \\
                  & UCT      & 1.80  & 2.01  & 2.19  & 2.50  & 2.90 & 3.83  \\

\noalign{\smallskip}\hline\noalign{\smallskip}

            & OBEP    & 2.62 & 6.14 & 11.73 & 14.99 & 16.65 & 17.40  \\
{${^3}P_2$} & OBEP$^*$& 2.62 & 6.14 & 11.73 & 14.99 & 16.65 & 17.39  \\
            & UCT$^*$ & 2.80 & 6.61 & 12.71 & 16.28 & 18.10 & 18.91  \\
            & UCT     & 2.57 & 6.00 & 11.32 & 14.18 & 15.37 & 15.07  \\

\noalign{\smallskip}\hline\noalign{\smallskip}

            & OBEP     & 0.11 & 0.34 & 0.77 & 1.04 & 1.10 & 0.52  \\
{${^3}F_2$} & OBEP$^*$ & 0.11 & 0.34 & 0.77 & 1.04 & 1.10 & 0.52  \\
            & UCT$^*$  & 0.11 & 0.34 & 0.77 & 1.05 & 1.13 & 0.64  \\
            & UCT      & 0.11 & 0.34 & 0.75 & 1.00 & 1.03 & 0.41  \\

\noalign{\smallskip}\hline\noalign{\smallskip}

                  & OBEP     & -0.86 & -1.82 & -2.84 & -3.05 & -2.85 & -2.02  \\
{$\varepsilon_2$} & OBEP$^*$ & -0.86 & -1.82 & -2.84 & -3.05 & -2.85 & -2.02  \\
                  & UCT$^*$  & -0.87 & -1.83 & -2.82 & -2.99 & -2.75 & -1.88  \\
                  & UCT      & -0.86 & -1.83 & -2.84 & -3.05 & -2.89 & -2.18  \\

\noalign{\smallskip}\hline\noalign{\smallskip}

\end{tabular}
\end{center}
\end{table*}
\begin{figure*}[!ht]
\begin{center}
  \includegraphics[scale=0.82]{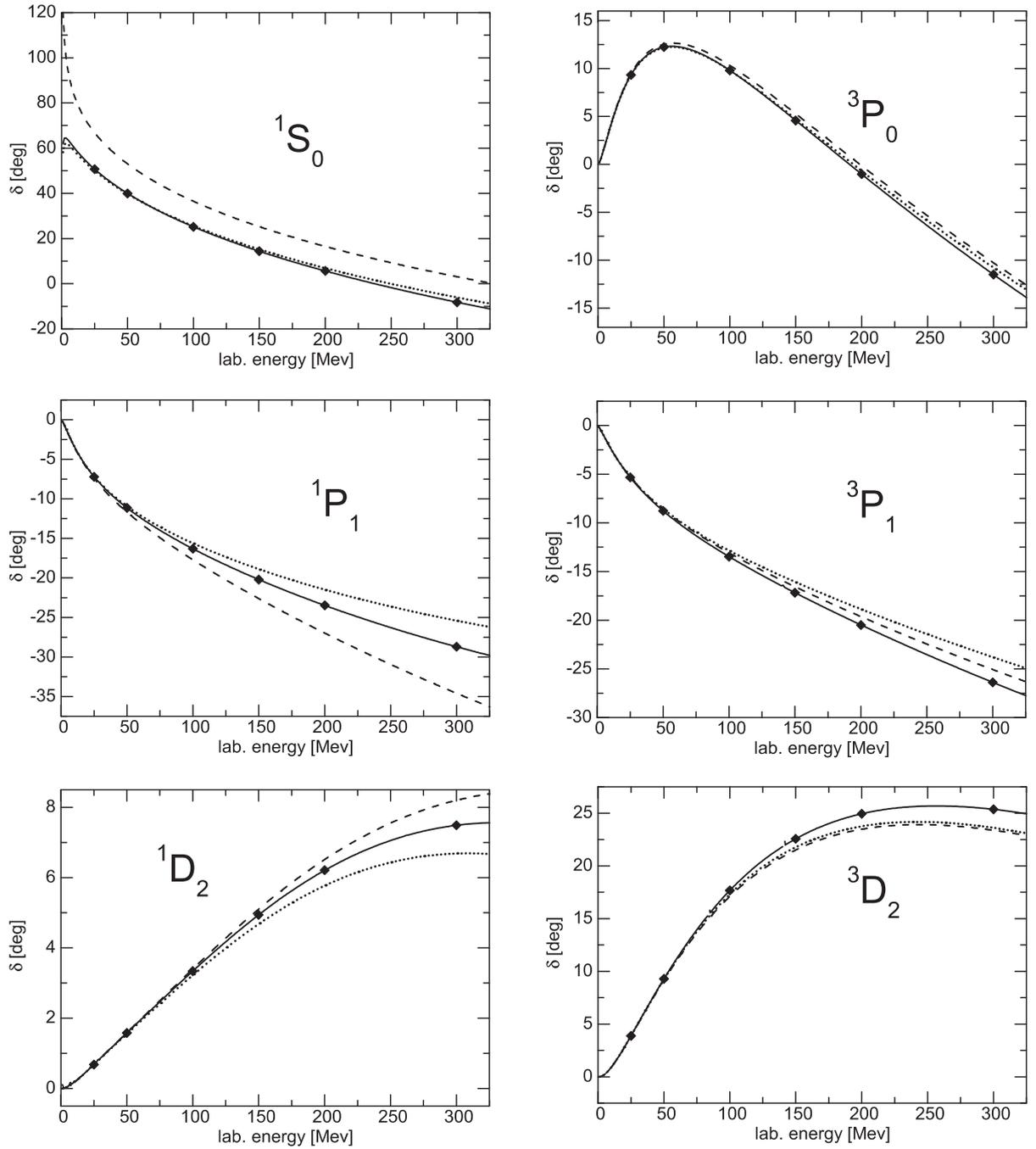}
\end{center}
\caption{Neutron-proton phase  parameters for the uncoupled
partial waves, plotted versus the nucleon kinetic energy in the lab. system.
Dashed[solid] curves calculated with Potential B parameters (Table \ref{UCT_tab:1})  by solving Eqs. \eqref{UCT_Eq.56}[\eqref{UCT_Eq.61}]. Dotted represent the solutions of Eqs. \eqref{UCT_Eq.56} with UCT parameters (Table \ref{UCT_tab:1}). The rhombs show original OBEP results (see Table 2).}
\label{UCT_fig:2}
\end{figure*}
\begin{multline*}
\left\langle {\vec p^{\, \prime} \,\mu'_1 \mu' _2 } \right|   v^{UCT}_{\rm {v}} \left| {\vec p\,\mu _1 \mu _2 } \right\rangle
= - \frac{F^2_{\rm v}[(p'-p)^2]}{\left({p' - p}\right)^2  - m_{\rm{v}}^2 } \\
\shoveleft{ \times \left\{ \bar u (\vec p^{\, \prime}) \left[ \left( g_{\rm{v}} + f_{\rm{v}} \right) \gamma _{\nu} - \frac{f_{\rm{v}}}{2m} \left( p' + p \right)_{\nu } \right. \right.}\\
\shoveright{ \left. - \frac{f_{\rm{v}}}{2m} (E_{\vec p'} - E_{\vec p})[\gamma_0 \gamma_{\nu}-g_{0 \nu}] \right] u\left(\vec p\right) }  \\
\shoveleft{\times \bar u\left(- \vec p^{\, \prime} \right) \left[ \left({g_{\rm{v}} + f_{\rm{v}}} \right) \gamma ^\nu - \frac{f_{\rm{v}}}{2m} \overline{\left( p' + p \right)}^{\nu} \right.}  \\
 \left. - \frac{f_{\rm{v}}}{2m} (E_{\vec p'} - E_{\vec p})[\gamma^0 \gamma^{\nu}-g^{0 \nu}] \right]u(- \vec p)
\end{multline*}
\begin{multline}
- \frac{{f_{\rm v}}^2}{4m^2} (E_{p'}-E_p)^2 \bar u(\vec p^{\, \prime})[\gamma_0 \gamma_{\nu}-g_{0 \nu}]u(\vec p) \\
\left. \phantom{\frac{f_{\rm{v}}}{2m}} \times \bar u(-\vec p^{\,
\prime})[\gamma^0 \gamma^{\nu}-g^{0 \nu}]u(-\vec p) \right\},
\label{UCT_Eq.59}
\end{multline}
where $\overline{( p' + p)}^{\nu} = (E_{\vec p'} + E_{\vec p},
-(\vec p' + \vec p))$.

At first sight, such a regularization can be achieved via a simple substitution $g_b \rightarrow g_b F_b(p',p)$ with some cutoff functions $F_b(p',p)$ depending on the 4--momenta $p'$ and~$p$. However, the principal moment is to satisfy the requirement \eqref{UCT_Eq.22} for the Hamiltonian invariant under space
inversion, time reversal and charge conjugation. In this context, let us remind that the baryon--nucleon--nucleon form factors are expressed through the matrix elements $\langle p' | j_b(0) | p \rangle$ of the corresponding baryon current density $j_b(x)$ at $x=0$ between  physical(clothed) one--nucleon states \cite{Gas}. Such matrix elements might be evaluated in terms of the cutoffs $F_b(p',p)$ using some idea from \cite{SheShi00} (cf. the clothed particle representation of a current therein).
\begin{figure*}[t]
\begin{center}
  \includegraphics[scale=0.82]{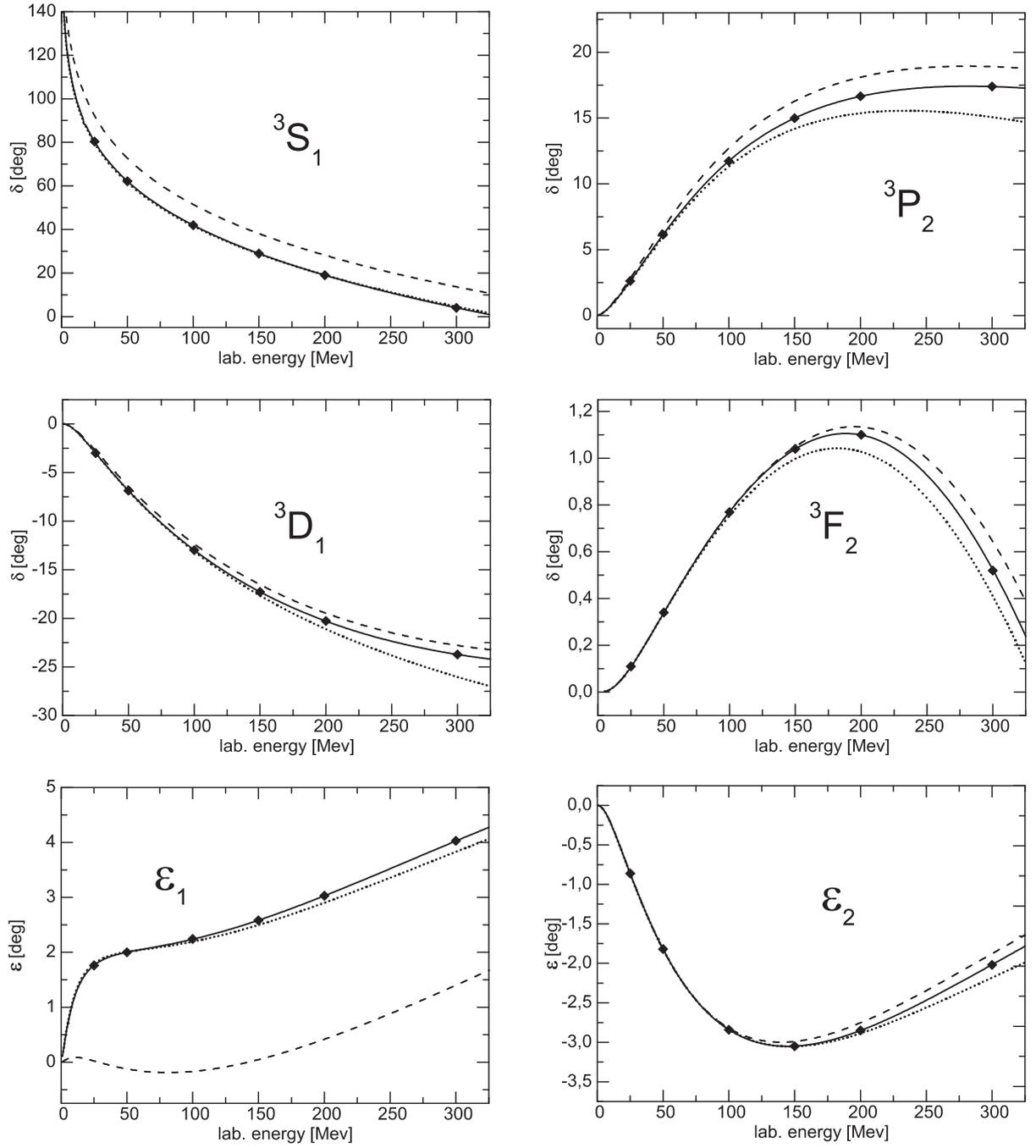}
\end{center}

\caption{The same in Fig. 2 but for the coupled waves.}
\label{UCT_fig:3}
\end{figure*}

Replacing in equations \eqref{UCT_Eq.57}--\eqref{UCT_Eq.59}
\begin{equation*}
\frac {1}{(p' - p)^2  - m_b^2} \, F^2_b[(p'-p)^2]
\end{equation*}
by
\begin{equation*}
\frac{-1}{(\vec p' - \vec p)^2 + m_b^2} \, F^2_b[-(\vec p'- \vec p)^2]
\end{equation*}
and neglecting the tensor-tensor term
\begin{multline}
\frac{{f_{\rm v}}^2}{4m^2}(E_{p'}-E_p)^2  \\
\times \bar u(\vec p^{\, \prime})[\gamma_0 \gamma_{\nu}-g_{0 \nu}]u(\vec
p) \bar u(-\vec p^{\, \prime})[\gamma^0 \gamma^{\nu}-g^{0
\nu}]u(-\vec p)
\label{UCT_Eq.60}
\end{multline}
in \eqref{UCT_Eq.59}, we obtain approximate expressions that with the
common factor
\begin{equation*}
(2\pi )^{-3} m^2/E_{p'}E_{p}
\end{equation*}
instead of
\begin{equation*}
(2\pi)^{-3}m/\sqrt{E_{p'} E_{p}}
\end{equation*}
are equivalent to Eqs. (E.21)--(E.23) from \cite{MachHolElst87}. Such an equivalence becomes
coincidence if in our formulae instead of the canonical
two-nucleon basis  $\left| {\vec p\,\mu _1 \mu _2 } \right\rangle
$ one uses the helicity basis as in \cite{MachHolElst87}.

In the context, we have considered the set of equations
\begin{multline}
{^B}R_{l'{\rm{ }}l}^{JST} (p',p) = {^B} V_{l'{\rm{ }}l}^{JST}(p',p) \\
+ m \sum \limits_{l''} \int \limits_0^\infty \frac{d q}{p^2 - q^2} \left\{ q^2\, {^B}V_{l'l''}^{JST}(p',q) {^B}R_{l''l}^{JST}(q,p) \right. \\
\left. -  p^2\,{^B}V_{l'l''}^{JST}(p',p) {^B}R_{l''l}^{JST}(p,p)
\right\} ,
\label{UCT_Eq.61}
\end{multline}
where the superscript $B$ refers to the partial matrix elements of
the potential $B$ defined in \cite{Mach89} with the just mentioned
interchange of the bases.

Our calculations of the $R$ matrices that meet the equations
\eqref{UCT_Eq.56} and \eqref{UCT_Eq.61} are twofold. On the one hand, we
will check reliability of our numerical procedure (in particular,
its code). On the other hand, we would like to show similarities
and discrepancies between our results and those by the Bonn group
both on the energy shell and beyond it. These results are depicted
in Figs. \ref{UCT_fig:2}--\ref{UCT_fig:3} and collected in Table \ref{UCT_tab:2}.

As seen in Figs. \ref{UCT_fig:2}--\ref{UCT_fig:3}, the most appreciable distinctions between
the UCT and OBEP curves take place for the phase shifts with the
lowest $l-$values. As the orbital angular momentum increases the
difference between the solid and da\-shed curves decreases. Such
features may be explained if one takes into account that the
approximations under consideration affect mainly high--momentum
components of the UCT quasipotentials (their behavior at "small"
distan\-ces). With the $l$--increase the influence of small
distances is suppressed by the centrifugal barrier repulsion.

Of course, it would be more instructive to compare the corresponding half--off--energy--shell $R$--matrices (see definition \eqref{UCT_Eq.55}). Their $p'$--dependencies not shown here have been prepared for a separate publication.
They are necessary to know when calculating the $\psi^{(\pm)}$ scattering states for a two--nucleon system. In the context, one should emphasize that hitherto we have explored the OBEP and UCT $R$--matrices in the c.m.s.,
where the both approaches yield most close results. It is not the case in those situations when the c.m.s. cannot be referred to everywhere (e.g., in the reactions $NN \rightarrow \gamma NN$ and $\gamma d \rightarrow pn$).
 In this respect our studies of the differences between UCT and OBE approaches are under way.

\section{Summary}
\label{UCT_sum}
The present work has been made to develop a consistent
field--theoretical approach in the theory of nucleon--nucleon
scattering. It has been shown that the method of UCT's, based upon
the notion of clothed particles, is proved to be appropriate in
achieving this purpose.

Using the unitary equivalence of the CPR to the BPR, we have seen
how in the approximation $K_I = K_I^{(2)}$ the extremely
complicated scattering problem in QFT can be reduced to the three--dimensional LS--type equation for the $T$--matrix in
momentum space.The equation kernel is given by the clothed
two--nucleon interaction of the class [2.2]. Such a conversation
becomes possible owing to the property of $ K_I^{(2)}$ to leave
the two--nucleon sector and its separate subsectors to be
invariant.

Special attention has been paid to the elimination of auxiliary
field components. We encounter such a necessity for interacting
vector and fermion fields when in accordance with the canonical
formalism the interaction Hamiltonian density embodies not  only a
scalar contribution but nonscalar terms too. It has proved (at
least, for the primary $\rho N$ and $\omega N$ couplings) that the
UCT method allows us to remove such noncovariant terms directly in
the Hamiltonian. To what extent this result will take place in
higher orders in coupling constants it will be a subject of further
explorations.




\end{document}